\def\mycmd{2}

\if\mycmd1
\documentclass[11pt,onecolumn, draftcls]{IEEEtran}
\else 
\documentclass[lettersize,journal]{IEEEtran}
\fi

\usepackage[latin9]{inputenc}
\usepackage{tabularx}
\usepackage{fontawesome}
\usepackage{threeparttable}

\usepackage{colortbl}    
\usepackage{pifont}     

\usepackage[british]{babel}
\usepackage{multicol}
\usepackage{array,ragged2e}
\usepackage{float}
\usepackage{mathtools}
\usepackage{amsmath}
\usepackage{amsthm}
\usepackage{amssymb}
\usepackage{graphicx}
\usepackage{wasysym}
\usepackage{setspace}
\usepackage{color}
\usepackage{bm}
\usepackage{cases}
\usepackage{makecell}
\usepackage{tikz}
\usepackage{flowchart}
\usepackage{amsfonts}
\usepackage{steinmetz}
\usetikzlibrary{matrix,shapes,arrows,positioning,chains}
\usepackage{lmodern,babel,adjustbox,booktabs,multirow}
\usepackage{makecell}
\usepackage{pbox}
\usepackage{epsfig}
\usepackage{subfigure}
\usepackage{tcolorbox}
\usepackage{enumitem}

\makeatletter
\newcommand{\multiline}[1]{%
  \begin{tabularx}{\dimexpr\linewidth-\ALG@thistlm}[t]{@{}X@{}}
    #1
  \end{tabularx}
}

\floatstyle{ruled}
\newfloat{algorithm}{tbp}{loa}
\providecommand{\algorithmname}{Algorithm}
\floatname{algorithm}{\protect\algorithmname}

\theoremstyle{plain}

\theoremstyle{plain}

\theoremstyle{plain}

\usepackage{epsfig}
\usepackage[caption=false,font=normalsize,labelfont=sf,textfont=sf]{subfig}
\usepackage{cite}
\usepackage{stfloats}
\usepackage{graphicx}
\usepackage{multirow}
\usepackage{array}
\usepackage{graphicx}
\usepackage{times}
\usepackage{algorithm}
\usepackage{algpseudocode}
\usepackage{nicefrac}
\theoremstyle{remark}

\makeatother

\algrenewcommand\algorithmicindent{1.0em}%
\providecommand{\lemmaname}{Lemma}
\providecommand{\propositionname}{Proposition}

\providecommand{\theoremname}{Theorem}
\providecommand{\theoremname}{Definition}
\newcommand{\rom}[1]{\uppercase\expandafter{\romannumeral #1\relax}}

\newcounter{problem}
\newcounter{save@equation}
\newcounter{save@problem}

\definecolor{lightergray}{gray}{0.8}
\definecolor{ForestGreen}{RGB}{34,139,34}  

\newcommand{\Xmark}{\textcolor{lightergray}{\ding{55}}}

\newcommand{\Problem}[1]{\bf{\mathdutchcal{P}#1}}
\DeclareMathAlphabet{\mathdutchcal}{U}{eus}{b}{n}
\DeclareMathAlphabet{\mathcal}{OMS}{cmsy}{m}{n}
\SetMathAlphabet{\mathcal}{bold}{OMS}{cmsy}{b}{n}


\numberwithin{save@problem}{subsection}
\numberwithin{save@equation}{subsection}

\allowdisplaybreaks[4]

\makeatletter
\renewcommand\paragraph[1]{%
    \vspace{.1cm}\noindent\textbf{#1}
}
\makeatother

\begin{document}
\title{$\alpha$-Fair Multistatic ISAC Beamforming\\for Multi-User MIMO-OFDM Systems\\via Riemannian Optimization}
\author{Hyeonho Noh,~\IEEEmembership{Member,~IEEE} and Jonggyu Jang,~\IEEEmembership{Member,~IEEE}
\thanks{Hyeonho Noh is with the Department of Information and Communication Engineering, Hanbat National University, Republic of Korea (e-mail: hhnoh@hanbat.ac.kr). 
    Jonggyu Jang is with the Department of Electronics Engineering, Chungnam National University, Daejeon 34134, Republic of Korea (e-mail: jgjang@cnu.ac.kr).
}
}

\maketitle
\begin{abstract}\label{abstract}
This paper proposes an $\alpha$-fair multistatic integrated sensing and communication (ISAC) framework for multi-user multi-input multi-output (MIMO)-orthogonal frequency division multiplexing (OFDM) systems, where communication users act as passive bistatic receivers to enable multistatic sensing. Unlike existing works that optimize aggregate sensing metrics and thus favor geometrically advantageous targets, we minimize the $\alpha$-fairness utility over per-target Cram\'{e}r--Rao lower bounds (CRLBs) subject to per-user minimum data rate and transmit power constraints. The resulting non-convex problem is solved via the Riemannian conjugate gradient (RCG) method with a smooth penalty reformulation. Simulation results validate the effectiveness of the proposed scheme in achieving a favorable sensing fairness--communication trade-off.
\end{abstract}


\section{Introduction}
\label{sec:introduction}
Integrated sensing and communication (ISAC) has recently emerged as a key enabling technology for next-generation wireless systems, offering a unified framework in which communication and sensing functionalities share the same spectrum, hardware, and waveform \cite{Wen25_CST, Zhu25_CST, Noh23_TVT}. Nevertheless, ISAC systems are subject to a fundamental asymmetry between the two functions: communication relies on one-way transmission from the base station to users, whereas sensing involves a two-way propagation process, incurring substantially higher path loss \cite{Noh25_arxiv_RIS}. To overcome this inherent disadvantage, sensing typically demands wider bandwidth and additional spatial degrees of freedom to achieve reliable and high-resolution parameter estimation. Yet dedicating these resources to sensing comes at the direct expense of communication performance, giving rise to an unavoidable sensing--communication trade-off that lies at the heart of ISAC system design \cite{Xiong23_TIT}.

To address these limitations, collaborative sensing approaches have been proposed, in which communication users or passive receivers capture target echoes as bistatic receivers, exploiting the spatial diversity of geographically distributed nodes to compensate for the path loss~\cite{Guo23_TSP, Bazzi25_JSAC, Perera26_TVT, Behdad24_TWC, Wang26_TCOM, Yang26_TWC, Yuan26_TWC}, as summarized in Table~\ref{tab:compare}. However, several of these works rely on restrictive system assumptions such as a single target~\cite{Guo23_TSP, Bazzi25_JSAC, Behdad24_TWC, Wang26_TCOM, Yang26_TWC, Yuan26_TWC}, a single passive receiver~\cite{Guo23_TSP, Bazzi25_JSAC, Perera26_TVT}, or an implicit waveform model~\cite{Guo23_TSP, Bazzi25_JSAC, Perera26_TVT, Behdad24_TWC, Wang26_TCOM, Yang26_TWC, Yuan26_TWC,Lou26_TCCN}. 
Moreover, most existing works are limited to specific fairness criteria: minimizing the sum CRLB, which implicitly prioritizes geometrically favorable targets, or employing a min--max formulation that focuses exclusively on the worst-case target~\cite{Lou26_TCCN}. However, a generalized framework that smoothly interpolates between these two extremes and enables flexible control over the fairness--efficiency trade-off in sensing performance remains absent in the literature.

In this paper, we propose an $\alpha$-fair multistatic ISAC framework for multi-user multi-input multi-output (MIMO)-orthogonal frequency division multiplexing (OFDM) systems, in which communication users act as passive bistatic receivers and jointly contribute sensing observations to enhance target parameter estimation. By tuning the fairness parameter $\alpha$, the proposed formulation spans a broad family of sensing objectives---from sum-optimal sensing that minimizes the aggregate CRLB, to min-max fair sensing that minimizes the worst-case CRLB. We formulate a joint transmit beamforming optimization problem that minimizes the $\alpha$-fairness utility over per-target CRLBs subject to per-user minimum data rate and total transmit power constraints. The resulting non-convex problem is solved via the Riemannian conjugate gradient (RCG) method on the complex sphere manifold, with per-user rate constraints enforced through a smooth penalty reformulation. Simulation results validate the effectiveness of the proposed scheme in achieving a favorable sensing fairness--communication trade-off.

\begin{table}
\centering
    \caption{Comparison of the proposed scheme against recent works}
    \begin{threeparttable}
    \adjustbox{width= \if 1\mycmd 0.6 \else 1.0 \fi \columnwidth}{
    \begin{tabular}{cccccc}
    \toprule
    \textbf{Ref.} & \textbf{Sensing Rx} & \textbf{Objective} & \textbf{Target} & \textbf{Waveform} & \textbf{Fairness} \\ 
    \midrule[\heavyrulewidth]\midrule[\heavyrulewidth]
    \arrayrulecolor{lightgray}
    \cite{Noh25_arxiv_RIS}         & BS            & Beamforming     & Single      & -  & \Xmark \\ \cline{1-6}
    \cite{Noh23_TVT, Noh25_arxiv_DCFNet}& BS            & Beamforming     & Multiple & OFDM & \Xmark \\ \cline{1-6}
    \cite{Guo23_TSP}         & Passive BS  & Waveform        & Single & -  & \Xmark \\ \cline{1-6}
    \cite{Bazzi25_JSAC}      & Passive BS  & Beamforming     & Single      & -  & \Xmark \\ \cline{1-6}
    \cite{Perera26_TVT}      & Passive BS  & Beamforming     & Multiple      & -  & Min-max \\ \cline{1-6}
    \cite{Lou26_TCCN}        & Multi-BSs    & Beamforming  & Multiple      & -  & Min-max \\ \cline{1-6}
    \cite{Behdad24_TWC}      & Multi-BSs    & Power Alloc.    & Single      & -  & \Xmark \\ \cline{1-6}
    \cite{Wang26_TCOM}        & Multi-users    & Beamforming     & Single       & -  & \Xmark \\ \cline{1-6}
    \cite{Yang26_TWC, Yuan26_TWC}       & Multi-BSs    & Beamforming     & Single       & -  & \Xmark \\ \cline{1-6}
    \textbf{Ours}            & BS, Multi-users   & Beamforming     & Multiple      & OFDM  & $\alpha$-fairness \\ 
    \arrayrulecolor{black}
    \bottomrule
    \end{tabular}
    }
    \end{threeparttable}
    \vspace{-10pt}
    \label{tab:compare}
\end{table}


\section{Scenario and Protocol}
\label{sec:system_model}

\begin{figure}[t]
    \centering
    \includegraphics[width=0.95\linewidth]{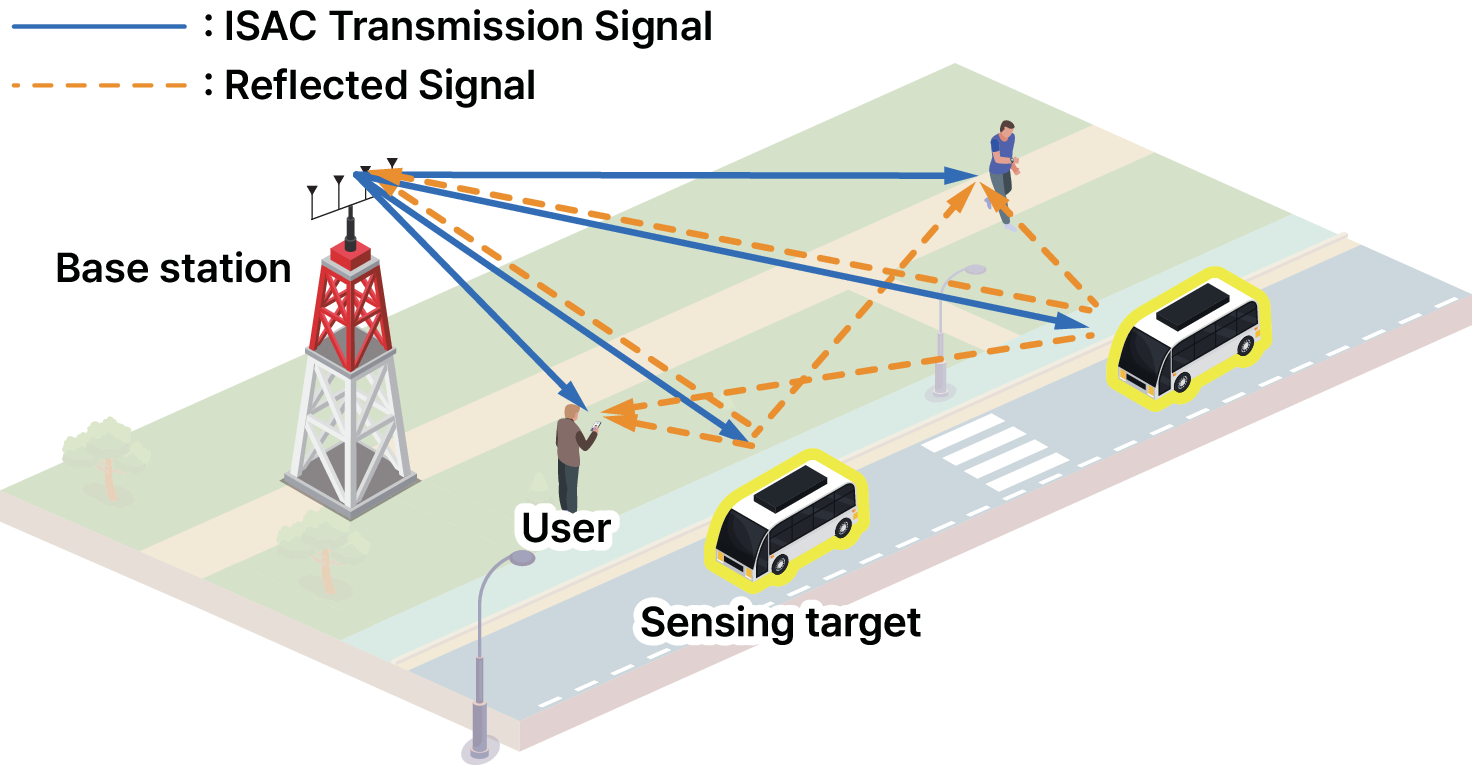}
    \vspace{-10pt}
    \caption{Illustration of the proposed multistatic ISAC scenario.}
    \label{fig:system_model}
    \vspace{-10pt}
\end{figure}

\subsection{System and Signal Model}
\label{subsec:system_model}
We consider a multistatic ISAC system with one base station (BS), $K$ communication users and $Q$ sensing targets, as shown in Fig.~\ref{fig:system_model}. 
The sets of communication users and sensing targets are defined by $\mathcal{K}=\{0,\ldots,K-1\}$ and $\mathcal{Q}=\{0,\ldots,Q-1\}$, respectively.
For the $i$-th subcarrier and $\mu$-th OFDM signal, the transmit data vector is defined as
\begin{align}
\mathbf d_i^{(\mu)}
= \sum_{k \in \mathcal{K}} \mathbf{v}_{k,i} c_{k,i}^{(\mu)} + \mathbf{w}_i s^{(\mu)},
\end{align}
where $\mathbf v_{k,i}, \mathbf w_{i} \in \mathbb{C}^{N_\mathrm{T}\times 1}$ denote the transmit beamforming vectors for the $k$-th communication user and the sensing signal on the $i$-th subcarrier, respectively. Moreover, $c_{k,i}^{(\mu)}$ and $s^{(\mu)}$ denote the communication symbol and the reference sensing symbol on the $i$-th subcarrier of the $\mu$-th OFDM signal.
The communication symbol $c_{k,i}^{(\mu)}$ and the sensing reference symbol $s^{(\mu)}$ are assumed 
to be independent and identically distributed (i.i.d.) with zero mean and unit variance. Furthermore, $s^{(\mu)}$ 
is assumed to be known at the receiver.
The corresponding OFDM transmit signal is given by
\begin{align} \label{eq:OFDM_signal}
\mathbf x^{(\mu)}(t)
=
\frac{1}{\sqrt{N_c}}
\sum_{i=0}^{N_c-1}
\mathbf d_i^{(\mu)}
e^{j2\pi i\Delta f t},
\end{align}
where $\Delta f = 1/T$ is the subcarrier spacing and $T$ is the OFDM symbol duration.
The up-converted signal is given as
$
\mathbf x^{(\mu)}_{\mathrm{RF}}(t)=\mathbf x^{(\mu)}(t)e^{j2\pi f_c t},
$
where $f_c$ denotes the carrier frequency.

Assuming perfect synchronization and CP removal, the received signal is represented in the frequency domain.
After down-converting and applying a discrete Fourier transform to the samples, the received signal from target $q$ on subcarrier $i$ at node $m \in \mathcal{K} \cup \{ \text{BS} \}$ is modeled as
\begin{align} \label{eq:rx_subcarrier}
\mathbf y_{q,m,i}^{(\mu)}
&=
\bar a_{q,m}
e^{-j2\pi i \bar \tau_{\text{BS},q,m}}
e^{j 2 \pi \bar f_{\text{D},\text{BS},q,m} \mu} \nonumber \\
& \qquad \cdot  \mathbf{a}_{\mathrm R}(\theta_{q,m}) \mathbf a_{\mathrm T}^{\text{H}}(\phi_{q})
\mathbf d_i^{(\mu)}
+
\mathbf n_{m,i}^{(\mu)},
\end{align}
where $\bar a_{q,m} = a_{q,m} e^{-j2 \pi f_\text{c} \tau_{\text{BS},q,m} }$, $a_{q,m}$ denotes the complex reflection coefficient of the $q$-th target at the $m$-th node, which incorporates the path loss, radar cross section, and a random reflection phase, $\bar f_{\text{D},\text{BS},q,m} = f_{\text{D},\text{BS},q,m} T_\text{sym}$ and $\bar \tau_{\text{BS},q,m} = \tau_{\text{BS},q,m} \Delta f$ are the normalized Doppler shift and delay of the target at $t=0$, respectively, $\phi_q$ and $\theta_{q,m}$ are the angle of departure (AoD) from the BS to target $q$ and angle of arrival (AoA) from target $q$ to node $m$, respectively,
$T_{\text{sym}}$ is the OFDM signal interval, $\mathbf n_{m,i}^{(\mu)} \sim \mathcal{CN}(\mathbf 0,\sigma_m^2\mathbf I)$ denotes the additive noise vector,
\begin{align}
     f_{\text{D},\text{BS},q,m} &= - \frac{v_{\text{BS},q,m}f_\text{c}}{c_0} (\cos(\phi_{q}) + \cos(\theta_{q,m}) ), \\
     \tau_{\text{BS},q,m} &= (d_{\text{BS},q} + d_{q,m})/c_0, 
\end{align}
$v_{\text{BS},q,m}$ is the radial velocity of target $q$ with respect to the BS--target--node $m$ path, and $d_{\text{BS},q}$ and $d_{q,m}$ denote the distances from the BS to target $q$ and from target $q$ to node $m$, respectively.

\subsection{Performance Metrics} 
The signal-to-interference-plus-noise ratio (SINR) of user $k$ on subcarrier $i$ is given by
\label{subsec:communication_SINR}
\begin{align}
\gamma_{k,i}
=
\frac{\big|(\mathbf h_{k,i})^{\mathrm H}\mathbf{v}_{k,i}\big|^2}
{\sum\limits_{\ell\in\mathcal K\setminus\{k\}}
\big|(\mathbf h_{k,i})^{\mathrm H}\mathbf{v}_{\ell,i}\big|^2
+
\big|(\mathbf h_{k,i})^{\mathrm H}\mathbf{w}_i\big|^2
+
\sigma_k^2 },
\end{align}
where $\mathbf{h}_{k,i} \in \mathbb{C}^{
N_\text{T} \times 1}$ is the communication channel vector from the BS to the $k$-th user on the $i$-th subcarrier and $\sigma_k^2$ is the noise variance. Denoting the bandwidth as $B$, we represent the data rate by $R_k = B \sum_{i=0}^{N_\text{c}-1} \log (1 + \gamma_{k,i})$.

We characterize the sensing performance of the proposed ISAC system in terms of the Fisher information matrix (FIM) and the CRLB for delay and Doppler estimation.
We denote FIM for $\boldsymbol{\eta}_{q,m}
=
\begin{bmatrix}
\bar \tau_{\mathrm{BS},q,m},
\bar f_{\mathrm D,\mathrm{BS},q,m}
\end{bmatrix}^{\mathsf T}$ as $
\mathbf J_{q,m,i}$. The explicit expressions of $\mathbf{J}_{q,m,i}$ are derived in Appendix~\ref{appendix:derivation_FIM}.

The multistatic sensing information for target $q$ is obtained by combining the monostatic sensing information at the BS and the bistatic sensing information from the users as
\begin{align}
\tilde{\mathbf J}_q
=
\sum_{m \in \mathcal{K} \cup \{ \text{BS} \} } \sum_{i=0}^{N_\text{c}-1} \mathbf J_{q,m,i},
\end{align}
where $\mathbf J_{q,\mathrm{BS},i}$ denotes the monostatic FIM at the BS and
$\mathbf J_{q,k,i}$ denotes the bistatic sensing FIM obtained from user $k$.

\subsection{$\alpha$-Fairness Multistatic Sensing Optimization}

To balance fairness across different targets, we adopt an $\alpha$-fairness utility \cite{Diakonikolas20_arxiv} to minimize the CRLBs
\begin{align}
    F_\alpha \left(\{\mathbf{v}_{k,i}\}, \{\mathbf{w}_i\} \right)
    =
    \sum_{q\in\mathcal Q} \dfrac{\left(\mathrm{tr}\!\left(\tilde{\mathbf J}_q^{-1}\right)\right)^{1+\alpha}}{1+\alpha},
\end{align}
where $\alpha \in [0,\infty)$.
Then, the joint beamforming problem is formulated as
\begin{subequations}
\label{eq:opt}
\begin{alignat}{3}
    & \Problem{1}: && 
    \min_{\{\mathbf{v}_{k,i}\}, \{\mathbf{w}_{i}\}}
    \quad
    && F_\alpha \left(\{\mathbf{v}_{k,i}\}, \{\mathbf{w}_i\} \right)
    \label{opt:alpha_fair_sensing}
    \\
    & && ~~~~~\text{s.t.}\quad
    && R_k \ge R_\text{min},\quad k \in \mathcal{K}, \label{subeq:opt1_rate}
    \\
    &&&&& \sum_{i=0}^{N_\mathrm{c}-1} (\sum_{k\in\mathcal K}
    \|\mathbf{v}_{k,i}\|^2 + \|\mathbf{w}_{i}\|^2)
    \leq P, \label{subeq:opt1_power}
\end{alignat}
\end{subequations}
where $R_\text{min}$ is the minimum data rate and $P$ is the total transmit power budget of the BS.

\section{Methodology}
In this section, we address the beamforming optimization problem. Although most existing works \cite{Yuan26_TWC,Lou26_TCCN,Yang26_TWC} adopt semidefinite relaxation (SDR) to transform the problem into a more tractable form, such approaches often suffer from performance degradation due to rank-one approximation. To remedy this issue, we adopt a RCG method that optimizes the beamforming vectors while satisfying the communication and power constraints.

\subsection{Solution for Beamforming Variables}
Instead of explicitly handling the nonconvex rate constraints, we incorporate their violations into the objective function through a squared penalty term.
This leads to a smooth unconstrained formulation over the complex sphere manifold induced by the total transmit power constraint.

\subsubsection{Penalty-Based Reformulation}

To penalize the violation of the rate constraint in \eqref{subeq:opt1_rate}, we define
\begin{align}
\phi_k(\{\mathbf v_{\ell,i}\},\{\mathbf{w}_i\})
=
\left[
R_{\min}
-
B \sum_{i=0}^{N_c-1}\log(1+\gamma_{k,i})
\right]_+,
\end{align}
where $[x]_+=\max(x,0)$.
Then, the beamforming problem is reformulated as
\begin{subequations}
\label{eq:opt2_penalty}
\begin{alignat}{3}
    & \Problem{2}: && 
    \min_{\{\mathbf{v}_{k,i}\}, \{\mathbf{w}_{i}\}}
    \quad
    && F_\alpha \left(\{\mathbf{v}_{k,i}\}, \{\mathbf{w}_i\} \right) + \frac{\rho}{2}\sum_{k\in\mathcal K}\phi_k^2,
    \label{subeq:opt2_obj}
    \\
    & && ~~~~~ \text{s.t.}\quad
    && \eqref{subeq:opt1_power},
\end{alignat}
\end{subequations}
where $\rho>0$ denotes the penalty parameter.
A larger $\rho$ imposes a stronger penalty on rate violations and drives the solution closer to the feasible region.

\subsubsection{Riemannian Conjugate Gradient Method}

We solve \eqref{eq:opt2_penalty} using the RCG method on the complex sphere manifold.
To express the problem compactly, we define the stacked beamforming vector at the $r$-th iteration as
\begin{align}
\mathbf z_i^{(r)}
&=
\big[
(\mathbf v_{0,i}^{(r)})^\text{T},\ldots,(\mathbf v_{K-1,i}^{(r)})^\text{T},
(\mathbf w_{i}^{(r)})^\text{T} \big]^\text{T}, \\
\mathbf z^{(r)}
&=
\big[
(\mathbf z_0^{(r)})^\text{T},\ldots,(\mathbf z_{N_c-1}^{(r)})^\text{T}
\big]^\text{T} .
\end{align}
With the effective objective 
$
\bar{F}_\alpha(\mathbf z)
=
F_\alpha(\mathbf z)
+
\frac{\rho}{2}\sum_{k\in\mathcal K}\phi_k^2(\mathbf z)
$,
the Euclidean gradient is given by
\begin{align}
\label{eq:euclidean_grad}
\nabla \bar{F}_\alpha(\mathbf{z})
=
\nabla F_\alpha(\mathbf{z})
-
\rho \sum_{k \in \mathcal{K}} \phi_k(\mathbf z)\nabla R_k,
\end{align}
whose explicit derivation is provided in Appendix~\ref{appendix:gradient}.

The feasible set is given by 
$
\mathcal M
=
\left\{
\mathbf z\in\mathbb C^n:\|\mathbf z\|^2=P
\right\},
$
which defines a complex sphere manifold. The tangent space at $\mathbf z^{(r)}$ is given by
\begin{align}
T_{\mathbf z^{(r)}}\mathcal M
=
\left\{
\boldsymbol\eta\in\mathbb C^n:
\langle\mathbf z^{(r)},\boldsymbol\eta\rangle_\mathbb{R}=0
\right\},
\end{align}
where $\langle \mathbf a, \mathbf b \rangle_\mathbb{R} = \Re\{\mathbf a^\text{H} \mathbf b\}$.
The Riemannian gradient is obtained by projecting the Euclidean gradient onto the tangent space:
\begin{align}
\nabla_{\mathcal M}\bar{F}_\alpha(\mathbf z^{(r)})
=
\nabla \bar{F}_\alpha(\mathbf z^{(r)})
-
\frac{
\langle\mathbf z^{(r)},\nabla \bar{F}_\alpha(\mathbf z^{(r)})\rangle_\mathbb{R}
}{P}\mathbf z^{(r)}.
\end{align}

Let $\boldsymbol\epsilon^{(r)}$ denote the search direction. 
Since $\boldsymbol\epsilon^{(r-1)} \in T_{\mathbf z^{(r-1)}}\mathcal M$ does not generally belong to $T_{\mathbf z^{(r)}}\mathcal M$, we project it onto the current tangent space. 
For the complex sphere manifold, the projection operator is given by
\begin{align}
\operatorname{Proj}_{\mathbf z^{(r)}}(\mathbf d)
=
\mathbf d
-
\frac{
\langle\mathbf {z}^{(r)},\mathbf{d}\rangle_\mathbb{R}
}{P}\mathbf z^{(r)},
\quad
\mathbf d \in \mathbb{C}^n.
\end{align}

The search direction is then updated as
\begin{align}
\boldsymbol\epsilon^{(r)}
=
- \nabla_{\mathcal M}\bar{F}_\alpha(\mathbf z^{(r)})
+
\beta^{(r)}
\operatorname{Proj}_{\mathbf z^{(r)}}
\big(\boldsymbol\epsilon^{(r-1)}\big),
\end{align}
where $\beta^{(r)}$ is the conjugate coefficient. 
We adopt the Polak--Ribi\`ere parameter as
\begin{align}
\beta^{(r)}
\hspace{-3pt}=\hspace{-3pt}
\frac{
\left\langle
\nabla_{\hspace{-3pt}\mathcal M}\bar{F}_\alpha(\mathbf z^{(r)}\hspace{-1pt}),
\nabla_{\hspace{-3pt}\mathcal M}\bar{F}_\alpha(\mathbf z^{(r)}\hspace{-1pt})
\hspace{-2pt}-\hspace{-2pt}
\operatorname{Proj}_{\mathbf z^{(r)}} \hspace{-3pt}
\big(
\nabla_{\hspace{-3pt}\mathcal M}\bar{F}_\alpha(\mathbf z^{(r-1)}\hspace{-1pt})
\big)
\right\rangle
}{
\left\langle
\nabla_{\hspace{-3pt}\mathcal M}\bar{F}_\alpha(\mathbf z^{(r-1)}\hspace{-1pt}),
\nabla_{\hspace{-3pt}\mathcal M}\bar{F}_\alpha(\mathbf z^{(r-1)}\hspace{-1pt})
\right\rangle
}.
\end{align}

A step size $\delta^{(r)}$ is determined via a line search, and the iterate is updated using the retraction
\begin{align}\label{eq:iteration}
\mathbf z^{(r+1)}
=
\mathcal R_{\mathbf z^{(r)}}\big(\delta^{(r)}\boldsymbol\epsilon^{(r)}\big)
=
\sqrt{P}\,
\frac{\mathbf z^{(r)}+\delta^{(r)}\boldsymbol\epsilon^{(r)}}{\|\mathbf z^{(r)}+\delta^{(r)}\boldsymbol\epsilon^{(r)}\|}.
\end{align}

The procedure \eqref{eq:iteration} is repeated until convergence.

\section{Simulation Results}

\begin{figure}[t]
    \centering
    \includegraphics[width=0.9\linewidth]{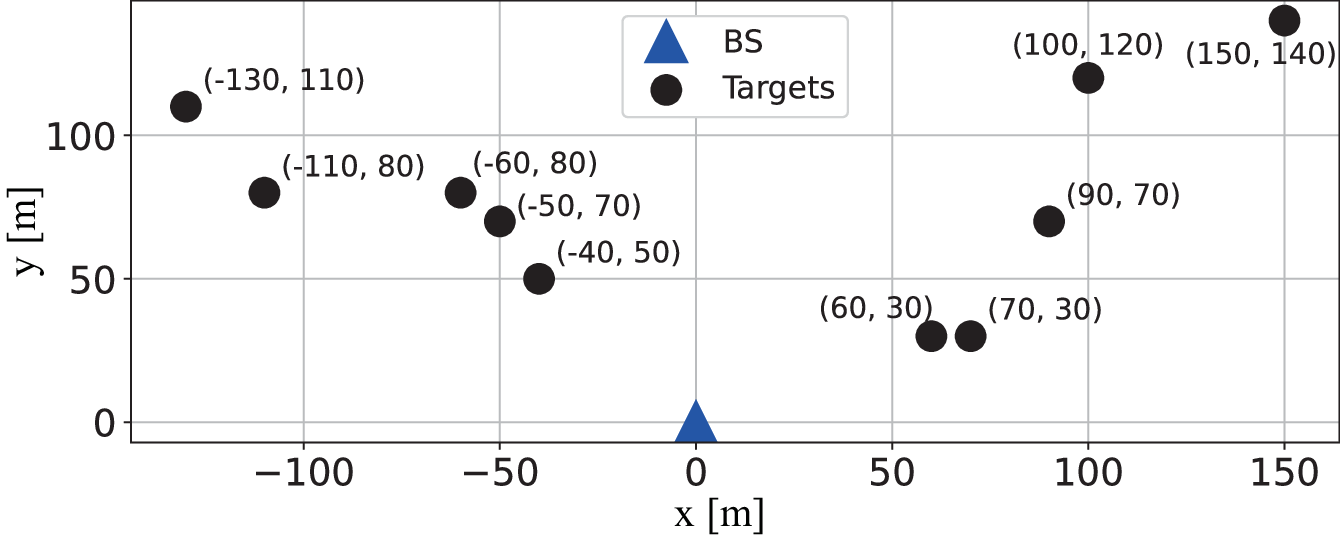}
    \vspace{-10pt}
    \caption{Position of the BS and sensing targets.}
    \label{fig:position}
    \vspace{-8pt}
\end{figure}

For simulations, the system parameters are chosen from the literature \cite{Noh25_arxiv_DCFNet}. The BS is equipped with $N_\text{T} = 20$ antennas and operates at a carrier frequency of $f_\text{c} = 28~\text{GHz}$ with a bandwidth of $B = 100~\text{MHz}$. The numbers of OFDM subcarriers and OFDM signals are $N_\text{c} = 2048$ and $N_\text{sym} = 64$, respectively, with a transmit power of $P = 30~\text{dBm}$. It is worth noting that the proposed method is not restricted to this specific parameter choice. The BS and 10 sensing targets are placed as shown in Fig.~\ref{fig:position} while 10 communication users are randomly distributed within a $30$~m radius of each target. The channel from the BS to each user is modeled based on 3GPP TR 38.901 \cite{3gpp38901}. The noise variance $\sigma_k^2$ is set to $-97$~dBm for all the users and the BS.

\subsection{Performance Analysis}


\begin{figure}[t]
    \centering
    \includegraphics[width=0.8\linewidth]{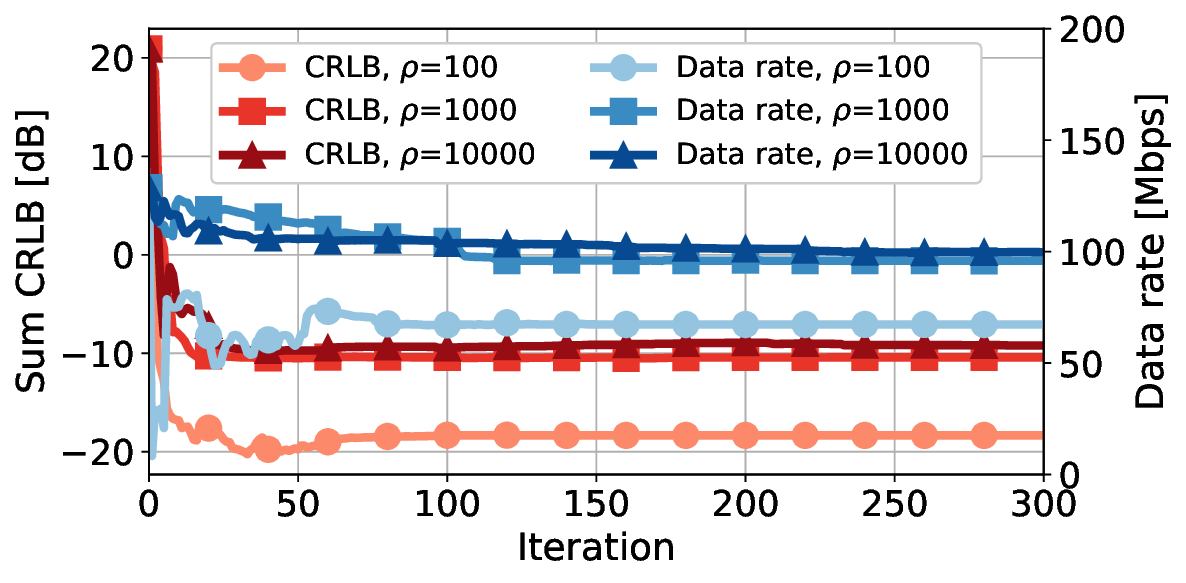}
    \vspace{-10pt}
    \caption{Convergence of sum CRLB and data rate for different penalty parameters $\rho$.}
    \label{fig:convergence}
    \vspace{-8pt}
\end{figure}

We first examine the convergence behavior of the proposed RCG algorithm with a minimum data rate of $100$~Mbps. As shown in Fig.~\ref{fig:convergence}, both the sum CRLB and the achievable data rate stabilize within approximately 50 iterations for all tested penalty parameters $\rho \in \{100, 1000, 10000\}$. A sufficiently large $\rho$ imposes a stronger penalty on rate violations, driving the solution toward the predefined data rate constraint, which validates our penalty-based reformulation.


\begin{figure}[t]
    \centering
    \includegraphics[width=\linewidth]{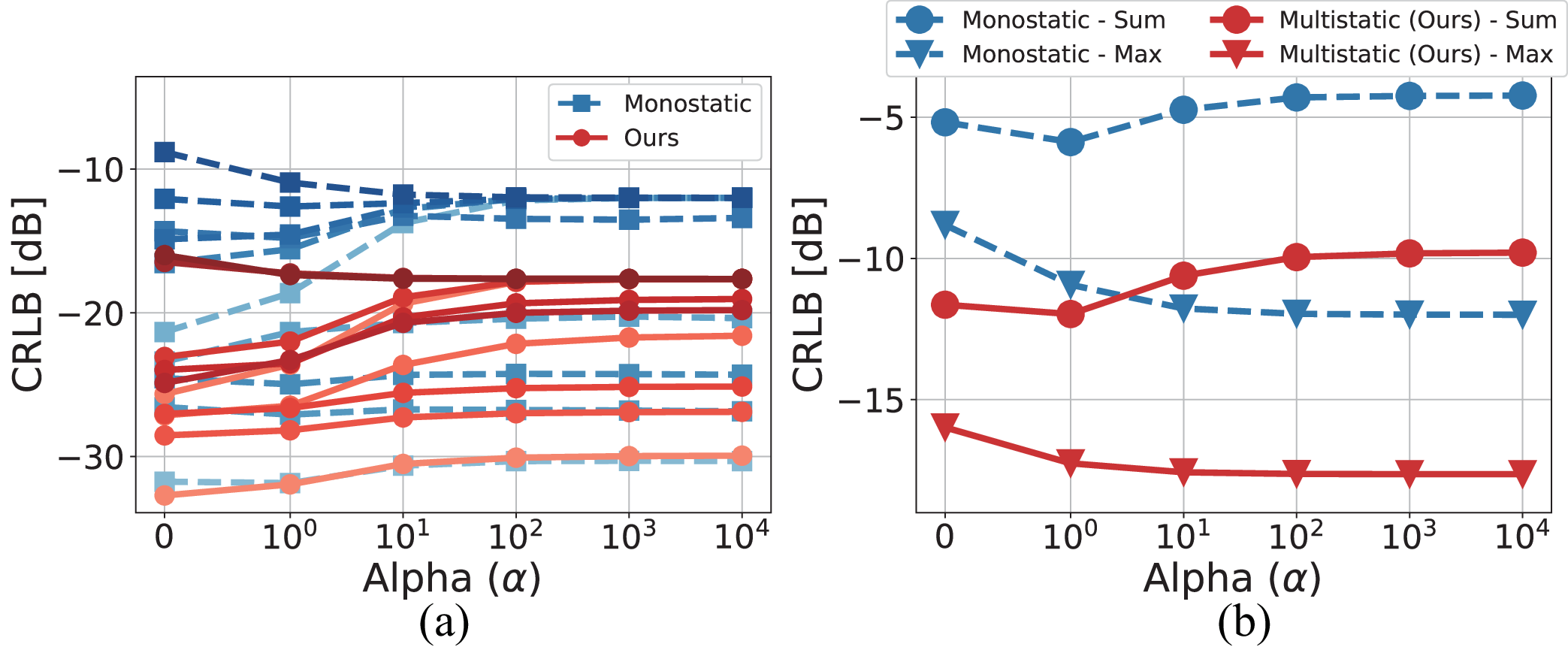}
    \vspace{-19pt}
    \caption{CRLB performance of 10 targets with different $\alpha$: (a) per-target CRLB and (b) sum and max CRLB.}
    \label{fig:alpha}
    \vspace{-8pt}
\end{figure}

Next, we investigate how the fairness parameter $\alpha$ shapes the sensing performance across targets. As shown in Fig.~\ref{fig:alpha}(a), the case $\alpha = 0$ reduces to the conventional sum-CRLB minimization, which predominantly favors geometrically advantageous targets and results in a wide disparity among per-target CRLBs. As $\alpha$ increases, the problem shifts toward fair target sensing, where the worst-case target receives progressively higher priority and the gap between the maximum and minimum CRLBs across targets shrinks. The aggregate metrics in Fig.~\ref{fig:alpha}(b) corroborate this trend: the sum CRLB rises moderately with $\alpha$, whereas the maximum CRLB drops, confirming that sensing fairness is effectively enforced.


\begin{figure}[t]
    \centering
    \includegraphics[width=0.8\linewidth]{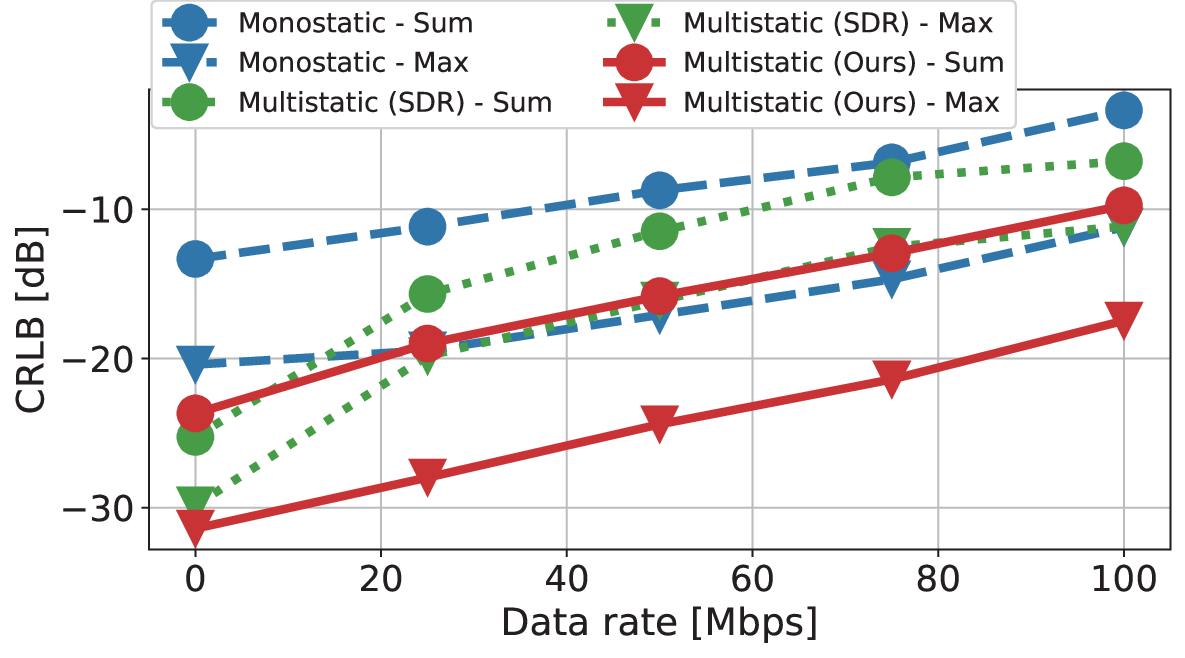}
    \vspace{-10pt}
    \caption{Trade-off between CRLB and data rate for monostatic, multistatic (SDR), and the proposed multistatic schemes.}
    \label{fig:tradeoff}
    \vspace{-8pt}
\end{figure}

The sensing--communication trade-off is further evaluated in Fig.~\ref{fig:tradeoff}, where three beamforming schemes are compared: monostatic, multistatic with successive convex approximation (SCA) and SDR \cite{Wang26_TCOM}, and our multistatic approach. Across all data rate requirements, the proposed scheme consistently attains lower sum and max CRLBs than both baselines. Two factors contribute to this gain: first, the multistatic architecture leverages the spatial diversity of geographically distributed receivers, thereby alleviating the severe two-way path loss that limits monostatic configurations; second, the RCG-based solver directly optimizes the beamforming vectors on the manifold without resorting to rank relaxation, which avoids the approximation loss inherent in SDR-based methods.


\begin{figure}[t]
    \centering
    \includegraphics[width=\linewidth]{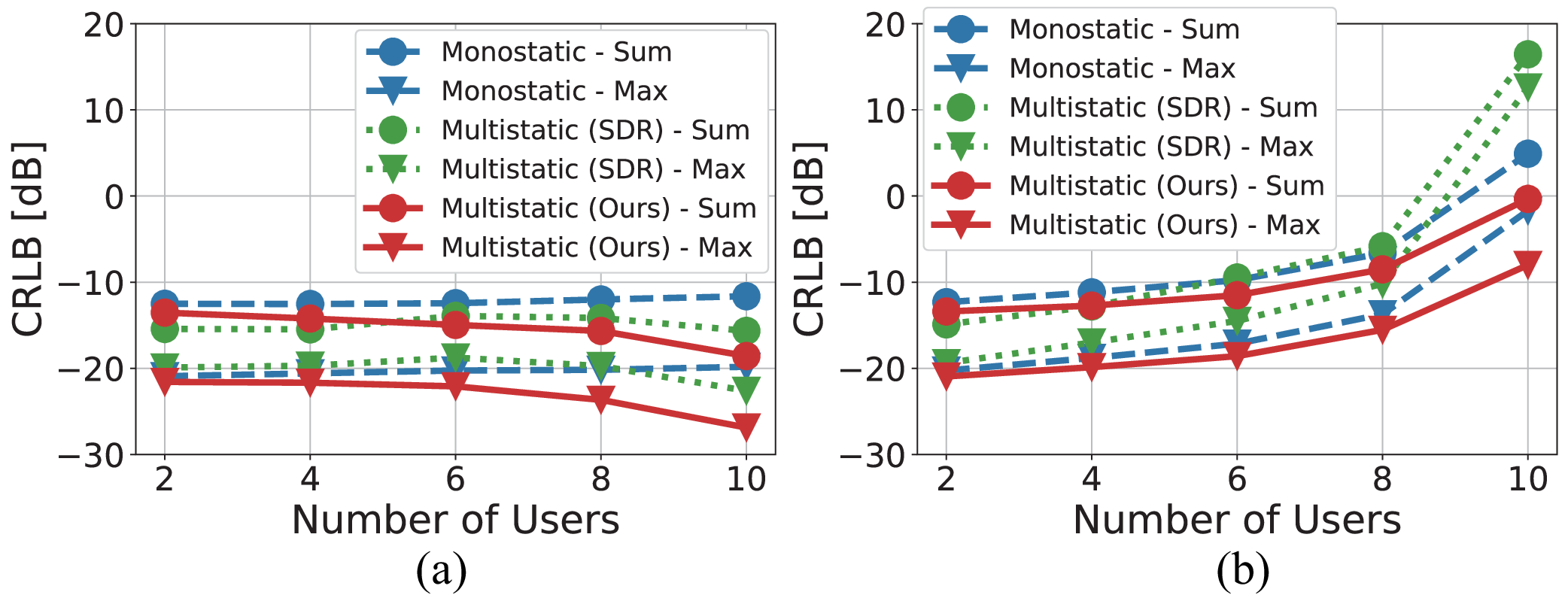}
    \vspace{-19pt}
    \caption{Sum and max CRLB versus the number of communication users for monostatic, multistatic (SDR), and the proposed multistatic schemes: (a) $R_\text{min} = 20~\text{Mbps}$ and (b) $R_\text{min} = 100~\text{Mbps}$.}
    \label{fig:crlb_user}
    \vspace{-8pt}
\end{figure}

Finally, we assess the scalability of the proposed framework by varying the number of communication users under two rate requirements. Under a relaxed constraint ($R_{\min} = 20~\text{Mbps}$), as depicted in Fig.~\ref{fig:crlb_user}(a), the proposed multistatic scheme sustains low CRLB values even as the number of users grows, since each additional user provides an extra bistatic observation that enriches the multistatic FIM. Both the monostatic and SDR-based baselines, by contrast, yield considerably higher CRLBs, with the monostatic scheme exhibiting slight degradation due to the communication constraints. When the rate requirement is tightened to $R_{\min} = 100~\text{Mbps}$ in Fig.~\ref{fig:crlb_user}(b), the CRLBs of all schemes increase with the number of users. Nonetheless, the proposed scheme maintains a substantial CRLB margin over both baselines, validating the robustness of the RCG-based multistatic sensing even under stringent communication demands.

\subsection{Computational Complexity}
The per-iteration cost of the proposed RCG algorithm is dominated by three operations. First, computing the sensing gradient $\nabla F_\alpha(\mathbf{z})$ requires evaluating the partial derivatives of $\xi^{(\mu)}_{q,m,i}$, which costs $\mathcal{O}(QK^2 N_c N_\text{T})$. Second, the Riemannian operations---gradient projection, conjugate direction update, and retraction---each cost $\mathcal{O}(N_c K N_\text{T})$. Third, the Armijo line search evaluates the objective $L$ times. Thus, the overall per-iteration complexity is $\mathcal{O}(LQK^2 N_c N_\text{T})$, which scales linearly in both $N_\text{T}$ and $N_c$, in contrast to the $\mathcal{O}(N_c(K{+}1)^{3.5}N_\text{T}^{3.5})$ cost of SDR-based interior-point methods \cite{Luo10_SPM}.

\section{Conclusion}
This paper proposed an $\alpha$-fair multistatic ISAC beamforming framework for multi-user MIMO-OFDM systems, where communication users serve as passive bistatic receivers. The joint beamforming problem was efficiently solved via the RCG method on the complex sphere manifold with a smooth penalty reformulation. Simulation results confirmed that the proposed scheme outperforms monostatic and SDR-based baselines in sensing accuracy while effectively balancing fairness and communication quality.

\begin{appendices}

\section{Derivation of FIM}
\label{appendix:derivation_FIM}
For notational convenience, we define the deterministic mean vector corresponding to the echo from target $q$ as
\begin{align}
\boldsymbol{\mu}_{q,m}(i,\mu)
=
\mathbf{c}_{q,m}(i,\mu)
e^{-j2\pi i \bar \tau_{\mathrm{BS},q,m}}
e^{j2\pi \bar f_{\mathrm{D},\mathrm{BS},q,m}\mu},
\end{align}
where the sensing waveform component is
\begin{align}
\mathbf{c}_{q,m}(i,\mu)
=
\frac{\bar a_{q,m}}{\sqrt{N_c}}
\mathbf a_\text{R}(\theta_{q,m})
\mathbf a_\text{T}^\text{H}(\phi_{q})
\mathbf{w}_i s^{(\mu)}.
\end{align}

For analytical tractability, the aggregate disturbance is approximated as spatially white interference-plus-noise. The resulting covariance matrix is modeled as $\mathbf R_m(i)=\sigma_{q,m,i}^2\mathbf I$, where
$
\sigma_{q,m,i}^2
=
\sigma_m^2
+
\sum_{\ell \in \mathcal{K} \setminus \{m\} }
\left|
\mathbf h_{m,i}^\text{H} \mathbf{v}_{\ell,i}
\right|^2 +
|\bar a_{q,m}|^2
\left(
\sum_{\ell\in\mathcal K}\|\mathbf{v}_{\ell,i}\|^2
+
\|\mathbf{w}_i\|^2
\right).
$

Since the received signal is complex Gaussian with parameter-dependent mean and covariance $\mathbf R_m$, the $(a,b)$-th entry of the FIM is given by
\begin{align}
&[\mathbf J_{q,m,i}]_{ab}
\\&=
2\Re
\Bigg\{
\sum_{\mu=0}^{N_{\text{sym}}-1}
\left(
\frac{\partial \boldsymbol{\mu}_{q,m}(i,\mu)}{\partial [\boldsymbol{\eta}_{q,m}]_a}
\right)^\text{H}
\mathbf R_m^{-1}(i)
\left(
\frac{\partial \boldsymbol{\mu}_{q,m}(i,\mu)}{\partial [\boldsymbol{\eta}_{q,m}]_b}
\right)
\Bigg\}. \nonumber
\end{align}

The multistatic FIM is then given by
\begin{align}
\label{eq:J_tilde_xi}
    [\tilde{\mathbf{J}}_q]_{ab} = \sum_{m \in \mathcal{K} \cup \{ \text{BS} \}} \sum_{i=0}^{N_\text{c}-1} \sum_{\mu=0}^{N_{\mathrm{sym}}-1} \kappa_{ab} \xi_{q,m,i},
\end{align}
where $\kappa_{11}=i^2$,
$\kappa_{22}=\mu^2$, $\kappa_{12}=\kappa_{21}=-i\mu$, and
$
\xi_{q,m,i}
=
\frac{8\pi^{2}|\bar{a}_{q,m}|^{2}}
     {N_c\,\sigma_{q,m,i}^{2}}
\left|
\mathbf{a}_\text{T}^\text{H}(\phi_{q}) \mathbf{w}_i \right|^{2}.
$

\section{Derivation of $\nabla \bar{F}_\alpha(\mathbf{z})$}
\label{appendix:gradient}

In this appendix, we derive the gradients of $R_k$ and $F_\alpha$ in~\eqref{eq:euclidean_grad} with respect to (w.r.t.) $\mathbf{v}_{\ell,i}^*$ and $\mathbf{w}_i^*$, respectively.

Applying the chain rule to $R_k$ yields
\begin{align}
\nabla_{\mathbf z^*} R_k
=
\sum_{i=0}^{N_c-1}
\frac{B}{1+\gamma_{k,i}}
\nabla_{\mathbf z^*}\gamma_{k,i}.
\end{align}

\paragraph{Gradient $\nabla_{\mathbf{v}_{\ell,i}^*} R_k$:}
We define $S_{k,i} = |\mathbf h_{k,i}^\text{H} \mathbf v_{k,i}|^2$ and $I_{k,i} = \sum_{\ell \in \mathcal{K} \backslash \{ k\} }|\mathbf h_{k,i}^\text{H} \mathbf v_{\ell,i}|^2
+ |\mathbf h_{k,i}^\text{H} \mathbf{w}_i|^2
+ \sigma_k^2$,
so that $\gamma_{k,i} = S_{k,i}/I_{k,i}$.
Applying the quotient rule to $\gamma_{k,i}$ with respect to $\mathbf{v}_{\ell,i}^*$ gives
\begin{align}
\nabla_{\mathbf v_{\ell,i}^*} R_k
=
\begin{cases}
\displaystyle
\frac{B}{1+\gamma_{k,i}}
\frac{1}{I_{k,i}}
\mathbf h_{k,i}\mathbf h_{k,i}^\text{H} \mathbf v_{k,i},
& \ell = k, \\[1em]
\displaystyle
-\frac{B}{1+\gamma_{k,i}}
\frac{S_{k,i}}{I_{k,i}^2}
\mathbf h_{k,i}\mathbf h_{k,i}^\text{H} \mathbf v_{\ell,i},
& \ell \neq k.
\end{cases}
\end{align}

\paragraph{Gradient $\nabla_{\mathbf{w}_{i}^*} R_k$:}
Since $\mathbf{w}_i$ appears only in $I_{k,i}$, an analogous derivation yields
\begin{align}
\nabla_{\mathbf{w}_i^*} R_k
=
-\frac{B}{1+\gamma_{k,i}}
\frac{S_{k,i}}{I_{k,i}^2}
\mathbf h_{k,i}\mathbf h_{k,i}^\text{H} \mathbf{w}_i.
\end{align}
\paragraph{Gradients $\nabla_{\mathbf{w}_{i}^*} F_\alpha$ and $\nabla_{\mathbf v_{\ell,i}^*} F_\alpha$:}
We now turn to the gradients of $\nabla F_\alpha$.
To this end, we find the gradient of $\nabla F_\alpha$ with respect to vector $\mathbf{x}$; then, we replace $\mathbf{x}$ by $\mathbf{w}_{i}$ and $\mathbf v_{\ell,i}$.
Using $\mathrm{d}\mathbf{X}^{-1} = -\mathbf{X}^{-1}(\mathrm{d}\mathbf{X})\mathbf{X}^{-1}$ and the chain rule, we have 
\begin{align}
\nabla_{\mathbf{x}^*} F_\alpha
    &= -\sum_{q \in \mathcal{Q}}
   \sum_{a\in\{1,2\}}
   \sum_{b\in\{1,2\}}
   [\mathbf{Q}_q]_{ba}\,
   \frac{\partial [\tilde{\mathbf{J}}_q]_{ab}}{\partial \mathbf{x}^*} \nonumber \\
    &= -\sum_{q \in \mathcal{Q}}
   \sum_{m \in \mathcal{K} \cup \{ \text{BS} \}}
   \sum_{i=0}^{N_\text{c}-1}
   \sum_{\mu=0}^{N_{\mathrm{sym}}-1}
   \Bigl(
     [\mathbf{Q}_q]_{11}\,i^2
     \nonumber \\
     &\quad -2[\mathbf{Q}_q]_{12}\,i\mu     
     +[\mathbf{Q}_q]_{22}\,\mu^2
   \Bigr)
   \frac{\partial \xi_{q,m,i}}{\partial \mathbf{x}^*},
\label{eq:dFalpha_final}
\end{align}
where $\mathbf{Q}_q =
\Bigl(\mathrm{tr}(\,\tilde{\mathbf{J}}_q^{-1})\Bigr)^{\alpha}\,
\tilde{\mathbf{J}}_q^{-2}$.

\paragraph{Derivatives $\nicefrac{\partial \xi_{q,m,i}}{\partial \mathbf{w}_i^*}$:}
Since both the numerator and denominator of $\xi_{q,m,i}$ depend on $\mathbf{w}_i$, the quotient rule gives
\begin{align}
\frac{\partial \xi_{q,m,i}}{\partial \mathbf{w}_i^*}
\hspace{-2pt}=\hspace{-2pt}
\left(\frac{8\pi^2}{N_c}
\,\mathbf{a}_\text{T}(\phi_{q})
\mathbf{a}_\text{T}^\text{H}(\phi_{q})
- \xi_{q,m,i} \mathbf{I} \right)
\frac{|\bar{a}_{q,m}|^2}{\sigma_{q,m,i}^{2}}
\mathbf{w}_i.
\label{eq:dxi_w}
\end{align}

\paragraph{Derivatives $\nicefrac{\partial \xi_{q,m,i}}{\partial \mathbf{v}_{\ell,i}^*}$:}
Since $\mathbf{v}_{\ell,i}$ enters only through $\sigma_{q,m,i}^2$, we obtain
\begin{align}
\frac{\partial \xi_{q,m,i}}{\partial \mathbf{v}_{\ell,i}^*}
=
\begin{cases}
\displaystyle
-\frac{\xi_{q,m,i}}{\sigma_{q,m,i}^{2}}
|\bar{a}_{q,m}|^{2}\,\mathbf{v}_{\ell,i},
& \ell = m, \\[1em]
\displaystyle
-\frac{\xi_{q,m,i}}{\sigma_{q,m,i}^{2}}
\left(
\mathbf{h}_{m,i}\mathbf{h}_{m,i}^\text{H}
+|\bar{a}_{q,m}|^{2}\right)\,\mathbf{v}_{\ell,i}
,
& \ell \neq m.
\end{cases}
\label{eq:dxi_v}
\end{align}


\end{appendices}

\bibliographystyle{IEEEtran}
\bibliography{{IEEEabrv,bibtex}}

\end{document}